\documentclass[%
 reprint,
 amsmath,amssymb,
 aip,
 apl,
]{revtex4-1}

\usepackage{graphicx}
\usepackage{dcolumn}
\usepackage{bm}
\usepackage[utf8]{inputenc}
\usepackage[T1]{fontenc}
\usepackage{mathptmx}
\usepackage{etoolbox}
\usepackage[table]{xcolor}
\makeatletter
\def\@email#1#2{%
 \endgroup
 \patchcmd{\titleblock@produce}
 {\frontmatter@RRAPformat}
 {\frontmatter@RRAPformat{\produce@RRAP{*#1\href{mailto:#2}{#2}}}\frontmatter@RRAPformat}
 {}{}
}%
\makeatother
\begin{document}

\newcommand{\smc}{\mathrm{g}_{\uparrow \downarrow}}

\newcommand{\alphaeff}{\alpha_{\rm eff}}

\newcommand{\Hk}{H_{\rm k}}

\newcommand{\Meff}{M_{\rm eff}}

\newcommand{\Ms}{M_{\rm \textrm{s}}}

\newcommand{\tFM}{t_{\rm FM}}

\newcommand{\tHM}{t_{\rm HM}}

\newcommand{\tMnSn}{t_{\rm Mn_3Sn}}

\newcommand{\Ho}{\Delta H_{\rm 0}}

\newcommand{\Ks}{K_{\rm s}}

\newcommand{\SHC}{\sigma_{\rm SH}}

\newcommand{\sdl}{\lambda_{\rm Mn_3Sn}}

\newcommand{\Vdc}{V_{\rm dc}}

\newcommand{\VISHE}{V_{\rm ISHE}}

\newcommand{\Vsym}{V_{\rm sym}}

\newcommand{\Vasym}{V_{\rm asym}}

\newcommand{\DelH}{\Delta H}

\newcommand{\Hr}{H_{\rm \textrm{r}}}

\newcommand{\HR}{H_{\rm \textrm{R}}}

\newcommand{\jc}{\vec{J}_{\rm \textrm{C}}}

\newcommand{\js}{\vec{J}_{\rm S}}

\newcommand{\Lr}{\lambda_{\textrm{IREE}}}

\newcommand{\alphaR}{\alpha_{\textrm{R}}}

\newcommand{\TPSH}{\theta_\textrm{PSH}}

\newcommand{\SHA}{\theta_\textrm{SH}}

\newcommand{\IC}{I_{\rm C}}

\preprint{AIP/123-QED}

\title{Large spin Hall conductivity in epitaxial thin films of kagome antiferromagnet Mn$_3$Sn  at room temperature}

\author{Himanshu Bangar}
 \affiliation{Department of Physics, Indian Institute of Technology Delhi, Hauz Khas, New Delhi 110016, India.}
 
\author{Kacho Imtiyaz Ali Khan}
\affiliation{Department of Physics, Indian Institute of Technology Delhi, Hauz Khas, New Delhi 110016, India.}

\author{Akash Kumar}
\affiliation{Department of Physics, University of Gothenburg, 412 96 Gothenburg, Sweden.}

\author{Niru Chowdhury}
\affiliation{Department of Physics, Indian Institute of Technology Delhi, Hauz Khas, New Delhi 110016, India.}
 
\author{Prasanta Kumar Muduli}
\affiliation{Department of Physics, Indian Institute of Technology Madras, Chennai, Tamil Nadu 600036, India.}
 
\author{Pranaba Kishor Muduli*}
 \email{muduli@physics.iitd.ac.in}
\affiliation{Department of Physics, Indian Institute of Technology Delhi, Hauz Khas, New Delhi 110016, India.}

\date{\today}

\begin{abstract}
Mn$_3$Sn is a non-collinear antiferromagnetic quantum material that exhibits a magnetic Weyl semimetallic state and has great potential for efficient memory devices.
High-quality epitaxial $c$-plane Mn$_3$Sn thin films have been grown on a sapphire substrate using a Ru seed layer. Using spin pumping induced inverse spin Hall effect measurements on $c$-plane epitaxial Mn$_3$Sn/Ni$_{80}$Fe$_{20}$, we measure spin-diffusion length ($\sdl$), and spin Hall conductivity ($\sigma_{\rm{SH}}$) of Mn$_3$Sn thin films: $\sdl=0.42\pm 0.04$~nm and $\sigma_{\rm{SH}}=-702~\hbar/ e~\Omega^{-1}$cm$^{-1}$. While $\sdl$ is consistent with earlier studies, $\SHC$ is an order of magnitude higher and of the opposite sign. The behavior is explained on the basis of excess Mn, which shifts the Fermi level in our films, leading to the observed behavior. Our findings demonstrate a technique for engineering $\sigma_{\rm{SH}}$ of Mn$_3$Sn films by employing Mn composition for functional spintronic devices.

\end{abstract}
\maketitle


Recently, there has been a huge amount of interest in quantum materials for the field of spintronics, which makes use of the electron's spin degree of freedom.~\cite{han2018quantum,vzutic2004spintronics}Spin-charge interconversion is important for the application of spintronics. Spin current can be generated from the charge current by a mechanism such as spin Hall effect (SHE), which is typically observed in non-magnetic heavy metals. However, recently non-collinear antiferromagnetic materials have gained attention as potential spin Hall materials. This is driven by (1) theoretical studies that predict large intrinsic spin Hall conductivity ($\SHC$),~\cite{zhang2018spin,busch2021spin} (2) the experimental observation of large anomalous Hall effect~\cite{nakatsuji2015large}, and (3) the observation of un-conventional spin-orbit torque~\cite{kondou2021giant} in these materials. 
Mn$_3$Sn is one example of non-collinear antiferromagnet that exhibits exotic properties such as anomalous Hall effect,~\cite{nakatsuji2015large} anomalous Nernst effect,~\cite{ikhlas2017large} and magneto-optic Kerr effect,~\cite{higo2018large} despite having nearly zero magnetization. These exotic properties originate from the Berry curvature associated with Weyl points near the Fermi energy.~\cite{chen2014anomalous,kubler2014non} More recently, new phenomena such as magnetic spin Hall effect,~\cite{kimata2019magnetic} chiral domain walls~\cite{li2019chiral} as well as spin-orbit torque induced chiral-spin rotation~\cite{takeuchi2021chiral} has been demonstrated in Mn$_3$Sn. These reports suggest that Mn$_3$Sn is a promising material for antiferromagnetic spintronics; a rapidly developing field that offers several advantages such as zero stray field, robustness against magnetic field perturbation, and ultrafast THz dynamics.~\cite{jungwirth2016antiferromagnetic,baltz2018antiferromagnetic} 

Theoretical works predict an intrinsic $\SHC$ in Mn$_3$Sn due to it's non-collinear magnetic structure.~\cite{zhang2018spin,busch2021spin, zhang2017prb} Only a few works are reported on the measurement of SHE in Mn$_3$Sn. The $\SHC$ of polycrystalline Mn$_3$Sn was recently reported to be $\sigma_{\rm{SH}}\sim 47~\hbar/ e~\Omega^{-1}$cm$^{-1}$ based on non-local spin transport experiments.~\cite{muduli2019evaluation} 
Yu \textit{et al.}~\cite{yu2021large} reported a spin Hall angle ($\SHA$) of 0.144 using the spin pumping induced inverse spin Hall effect (ISHE) approach, which is greater than that of Ta. To realize the full potential of Mn$_3$Sn for spintronics applications, it is important to estimate $\SHC$ as well as other important parameters like spin mixing conductance ($\smc$), and spin diffusion length ($\sdl$) in epitaxial thin films. Recently, Yoon $et al.$ reported a drastic change in transport properties with the crystalline orientation of Mn$_3$Sn thin films~\cite{yoon2019crystal}. Theoretical studies also predicts strong anisotropic $\SHA$ for Mn$_3$Sn.~\cite{zhang2017prb} Consequently, it is anticipated that the values of $\smc$, $\sdl$, and $\sigma_{\rm SH}$ will depend on the orientation of Mn$_3$Sn thin films.

In this work, we present direct measurement of $\SHC$ in high quality $c$-plane oriented Mn$_3$Sn thin films. We grow $c$-plane oriented Mn$_3$Sn on Al$_2$O$_3$ substrate using Ru as the seed layer. We determine the $\SHC$ by employing spin pumping driven ISHE measurements on \textit{epi-}Mn$_3$Sn/Ni$_{80}$Fe$_{20}$ bilayers. In contrast to previously reported investigations, $\SHC$ of our epitaxial Mn$_3$Sn thin films is an order of magnitude greater, while $\sdl$ is comparable. We also report $\smc$ of this system to be (1.54 $\pm$ 0.27) $\times 10^{19}~\rm{m}^{-2}$, which is also an order of magnitude higher compared to Py/Ta bilayers.~\cite{kumar2018large,deorani2013role} The sign of $\SHC$ is found to be negative, which can be explained by a shift of the Fermi level caused by a slight excess of Mn in our films.

\begin{figure*}[t]
\includegraphics[width=17cm]{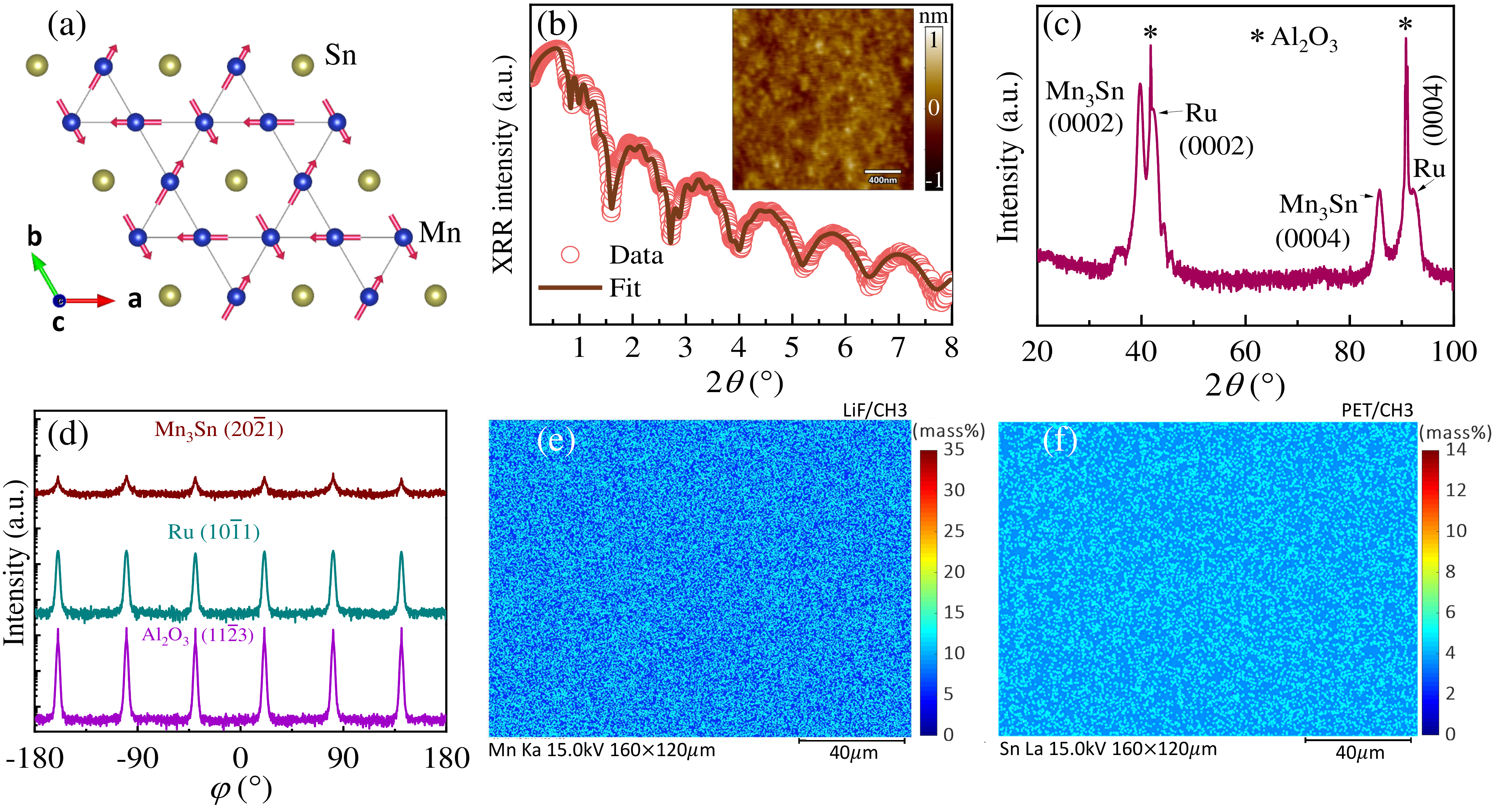}
\caption{\label{fig:1} (a) The magnetic and crystal structure of c-plane Mn$_3$Sn, revealing the kagome lattice of Mn magnetic moments. (b) Measured X-ray reflectivity (red circles) for the Ru(7 nm)/Mn$_3$Sn(30 nm)/AlO$_x$(2 nm) thin film and corresponding theoretical fit (black solid line). Inset: the surface topography of the same sample measured using AFM. The scan area is $2 ~\mu $m$ ~\times ~2 ~\mu $m$ $. (c) X-ray diffraction (XRD) spectra for the Ru(7 nm)/Mn$_3$Sn(30 nm)/AlO$_x$(2 nm) thin film. (d) The corresponding $\varphi$-scan for Al$_2$O$_3$ (11$\bar{2}$3) (violet), Ru (10$\bar{1}$1) (cyan) and Mn$_3$Sn (20$\bar{2}$1) (wine). Elements distribution maps for (e) Mn and (f) Sn measured using electron probe microanalyzer (EPMA).}
\end{figure*}

Mn$_3$Sn has a hexagonal crystal structure, for which the (0001)-plane or the $c$-plane has a kagome lattice as shown in Fig.~\ref{fig:1}(a). In order to grow the $c$-plane Mn$_3$Sn, we choose $c$-plane sapphire ($c$-Al$_2$O$_3$) substrate and a 7~nm thick Ru-seed layer. The films were grown using AJA Orion 8 magnetron sputtering system with a base pressure, better than $5\times10^{-8}$~Torr. 
First, the Ru seed layer was deposited at 400 $^{\circ}$C and annealed at the same temperature for 10 minutes. Then the sample was allowed to cool to 100 $^{\circ}$C, at which Mn$_3$Sn was deposited by co-sputtering Mn and Sn targets. The Mn was deposited at a growth rate of 0.37 \AA/s by applying 60 W DC power while the Sn was deposited at a growth rate of 0.20 \AA/s by applying 40 W RF power. After the deposition of Mn$_3$Sn layer, the sample was annealed \textit{in-situ} at 600~$^{\circ}$C for an hour. Subsequently, the sample was allowed to cool to room temperature, after which Ni$_{80}$Fe$_{20}$(8 nm)/Al(2 nm) was deposited without breaking the vacuum. The Al layer is used as the capping layer and is fully oxidized after exposure to air. 
During the deposition,
we rotate the sample holder along its normal axis to ensure a
uniform composition and thickness. The structural characterization was performed using high resolution X-ray diffraction (HR-XRD), while the growth rate and interfacial roughness were determined via X-ray reflectivity (XRR) measurements using a PANalytical x-ray
diffractometer equipped with Cu K$_{\alpha}$ ($\lambda$ = 1.5406 \AA) source. The morphology and surface roughness was determined by atomic force microscopy (AFM) scans (Asylum Research, MFP-3D system). The AFM images were obtained in the tapping mode using Asylum Research Probes (AC240TS-R3) cantilevers. The SHIMADZU (EPMA -1720 HT) electron probe microanalyzer (EPMA) system was used to determine the composition.
A broadband ferromagnetic resonance (FMR) spectroscopy technique was used to characterize the magnetization dynamics of Mn$_3$Sn/Ni$_{80}$Fe$_{20}$ bilayers. Field modulation technique with lock-in based detection is employed to enhance the sensitivity of FMR measurements.~\cite{kumar2019influence} 
The excitation radio frequency (RF) was varied from 3 to 8 GHz. 
To detect the ISHE voltage, copper pads were pasted beneath the inverted sample following the method used in our earlier work by Kumar $et$ $al.$~\cite{kumar2018large}


Figure~\ref{fig:1}(b) shows the XRR measurement of the Al$_2$O$_3$/Ru(7 nm)/Mn$_3$Sn(30 nm)/AlO$_x$(2 nm) thin film. The thickness of Mn$_3$Sn is extracted from the XRR fit, from which we determine the growth rate to be  $\approx$~0.48 \AA/s.
We obtained a low roughness ($\approx$~0.28 nm) from the fit of the XRR data. The AFM surface morphology also showed a lower root mean square roughness of $\approx$~0.15 nm measured  over a scan area of $2 ~\mu $m$ ~\times ~2 ~\mu $m$ $ as shown in the inset. Figure \ref{fig:1}(c) shows the corresponding 2$\theta$-$\theta$ XRD pattern, showing the (0002)-peaks corresponding to both Mn$_3$Sn and Ru seed layer. We also observe several satellite
peaks (thickness fringes), indicating smooth interfaces and uniform film growth, which is also consistent with sharp Kiessig fringes observed in the XRR measurements [Fig. \ref{fig:1}(b)]. We only observed (0002) and (0004) Mn$_3$Sn peaks, indicating that a $c$-plane oriented epitaxial Mn$_3$Sn film has been achieved on the Ru seed layer. To determine the epitaxial relationship, we performed $\varphi$-scans for the peaks: Al$_2$O$_3$ (11$\bar{2}$3) (violet), Ru (10$\bar{1}$1) (cyan) and Mn$_3$Sn (20$\bar{2}$1) (wine) as presented in Fig.~\ref{fig:1}(d). The $\varphi$-scan clearly shows that a reflection appears periodically every 60$^{\circ}$ for Al$_2$O$_3$, Ru, and Mn$_3$Sn, indicating that we have obtained non-twinned, highly epitaxial films. The location of the peaks indicates that the epitaxial film is formed with a relationship of Mn$_3$Sn (0001)[20$\bar{2}$0] || Ru (0001)[10$\bar{1}$0] || Al$_2$O$_3$ (0001)[11$\bar{2}$0], which is similar to the work by S.~Kurdi \textit{et al.}~\cite{kurdi2020optimization}

In order to determine the composition and its distribution, we used an electron probe microanalyzer (EPMA) and energy dispersive X-ray (EDX) mapping on our samples. The composition of the film is estimated to be Mn$_{3.14\pm 0.03}$Sn$_{0.86\pm 0.01}$ using EDX mapping and Mn$_{3.12\pm 0.02}$Sn$_{0.88\pm 0.01}$ by quantitative analysis using EPMA system. Both measurements show the presence of excess Mn, which is known to be essential for the formation of the D0$_{19}$ Mn$_3$Sn.~\cite{higo2018anomalous} Figure~\ref{fig:1}(e) and (f) provide the elemental distribution maps of Mn and Sn atoms, respectively, showing that the elements are uniformly distributed. We measured the longitudinal resistivity of Ru(7 nm)/Mn$_3$Sn(30 nm) bilayer using four point probe method to be 570 $\mu \Omega$-cm.

\begin{figure*}[t]
\includegraphics[width=17cm]{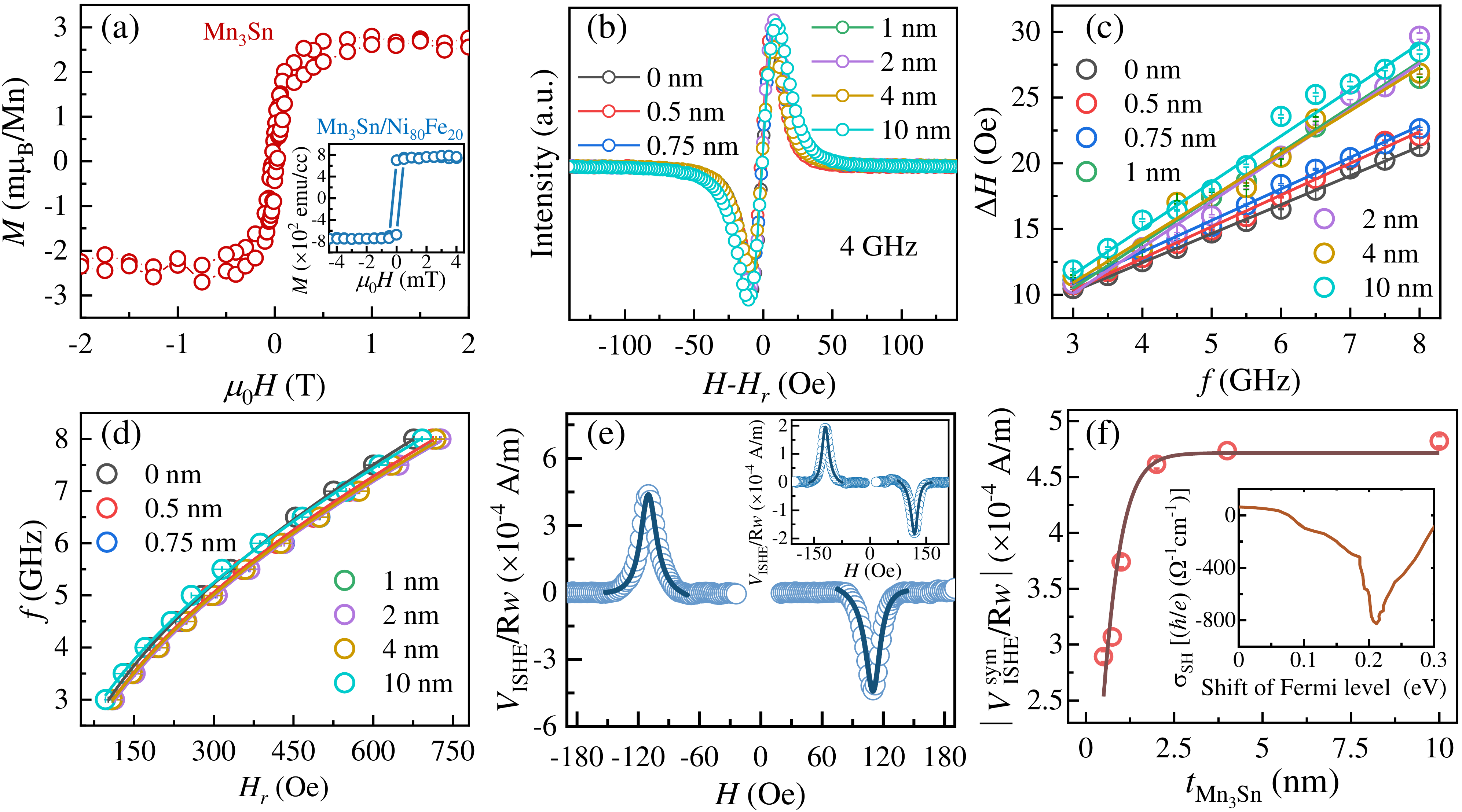}
\caption{\label{fig:2} (a) SQUID magnetization ($M$) curve of Mn$_3$Sn (30 nm) sample. Inset: Magnetization ($M$) curve of Mn$_3$Sn (30 nm)/Ni$_{80}$Fe$_{20}$ (8 nm) bilayer thin film. (b) The measured FMR signal at 4 GHz for samples (Al$_2$O$_3$/Ru/Mn$_3$Sn($\tMnSn$)/Ni$_{80}$Fe$_{20}$/AlO$_x$) having different thickness, $\tMnSn$ of Mn$_3$Sn near the resonance field ($\Hr$). (c) linewidth ($\Delta H$) vs. frequency ($f$) (open circle) and their corresponding fit (solid lines) with Eq. \ref{eq;deltah}. (d) $f$ vs. $\Hr$ curve for different $\tMnSn$ (open circle) and their corresponding fit (solid lines) with Eq. \ref{Kittel} (e) A representative ISHE signal for Mn$_3$Sn(10 nm)/Ni$_{80}$Fe$_{20}$(8 nm) sample. Inset: ISHE signal for a reference Ta(5 nm)/Ni$_{80}$Fe$_{20}$(8 nm) bilayer to establish the negative sign of spin Hall conductitivty ($\SHC$) in Mn$_3$Sn(10 nm) (f) The plot of the ratio $V_{\rm ISHE}^{\rm sym}/Rw$ as a function of Mn$_3$Sn thickness ($\tMnSn$) and its fit with Eq. \ref{Vfit} to determine $\sdl$ and $\SHA$. The inset depicts the predicted $\SHC$ based on Ref.~\cite{zhang2018spin}, when the Fermi level is moved upward with electron doping \textit{e.g.,} due to high Mn concentration.}
\end{figure*}

Magnetization ($M$) measurement performed using superconducting quantum interference device (SQUID) on a 30~nm thick Mn$_3$Sn thin film is shown in Fig.~\ref{fig:2}(a). It shows a weak magnetic moment of around 2 m$\mu _B $/Mn at room temperature, which is lower than polycrystalline films, as well as reported results on $c$-plane Mn$_3$Sn thin films.~\cite{higo2018anomalous,ikeda2020fabrication} Magnetization ($M$) measurement for Mn$_3$Sn/Ni$_{80}$Fe$_{20} $ (8~nm) bilayer is shown in the inset of Fig.~\ref{fig:2}(a), from which we determine saturation magnetization of Ni$_{80}$Fe$_{20}$ to be 750~emu/cc. In Fig. \ref{fig:2}(b), we show example FMR spectra of Al$_2$O$_3$/Ru/Mn$_3$Sn($\tMnSn$)/Ni$_{80}$Fe$_{20}$/AlO$_x$ for various thicknesses, $\tMnSn$ of Mn$_3$Sn measured at 4~GHz. The FMR data is fitted~\cite{celinski1997using} to extract the values of half-width at half maximum (HWHM) or linewidth ($\Delta H$) and resonance field ($H_{\rm r}$). 

The $\Delta H$ vs. frequency ($f$) dependence of all the samples are plotted in Fig.~\ref{fig:2}(c). The measured data is fitted using following equation:

\begin{equation}\label{eq;deltah}
\Delta H=\frac{2\pi\alpha_{\rm eff} f}{\gamma}+\Delta H_{\rm 0},
\end{equation}

where $\alphaeff$ is the effective Gilbert damping parameter, $\gamma = 1.85 \times 10^2 $ GHz/T is the gyromagnetic ratio, and $\Ho$ is the inhomogeneous line broadening. The first term on the right hand side is the viscous damping of magnetization motion, while the second term is due to magnetic inhomogeneity and sample imperfections of the FM layer.~\cite{farle1998ferromagnetic} The slope of the linear fit using Eq.~\ref{eq;deltah} is proportional to $\alphaeff$. As can be observed from Fig.~\ref{fig:2}(c), the slope of Mn$_3$Sn($\tMnSn$)/Ni$_{80}$Fe$_{20}$(8 nm) is higher compared to its corresponding reference Ni$_{80}$Fe$_{20}$(8 nm) sample, which indicates spin current being pumped from Ni$_{80}$Fe$_{20}$ into Mn$_3$Sn. Further, the low values of $\Ho$ ($<$ 3 Oe) from the fits indicate the high quality of our samples. Figure~\ref{fig:2}(d) shows the $f$ dependent $\Hr$ (open circles) for various $\tMnSn$ with their corresponding fit (solid line) using Kittel's equation:~\cite{kittel1948theory}

\begin{equation}\label{Kittel}
f=\frac{\gamma}{2\pi}[(H_{\rm r}+H_{\rm k})(H_{\rm r}+H_{\rm k}+4\pi M_{\rm eff})]^{1/2},
\end{equation}

where $\Hk$ is the uniaxial anisotropy field, and $\Meff$ is the effective saturation magnetization. The $\Meff$ is found to be comparable for all the Mn$_3$Sn($\tMnSn$)/Ni$_{80}$Fe$_{20}$ samples and in the range of $740-770$~emu/cc, which is close to measured saturation magnetization of 750~emu/cc (estimated from magnetization measurements), indicating negligible perpendicular anisotropy.

Next, we performed the Mn$_3$Sn thickness dependent ISHE measurements to determine the $\SHA$ of our $c-$plane Mn$_3$Sn films. Figure~\ref{fig:2}(e) shows a representative ISHE signal, $\VISHE$ (open circles) along with an ISHE signal measured from a Ta/Py/SiO$_2$ sample (inset). To eliminate contribution from self-induced ISHE and other rectification effects from Ni$_{80}$Fe$_{20}$\cite{gladii2019prb,conca2017lack,bangar2022}, the signal from a reference (Ni$_{80}$Fe$_{20}$) sample is subtracted from the measured data. The sign of signal obtained from Mn$_3$Sn agrees with Ta, indicating a negative sign of $\SHA$. The ISHE data ($V_{\rm ISHE}$) was fitted with a combination of symmetric ($V_{\rm ISHE}^{\rm sym}$) and asymmetric ($V_{\rm ISHE}^{\rm asym}$) components using equation:~\cite{deorani2014observation}

\begin{equation}\label{ViREE}
\begin{split}
\VISHE=V_{\rm ISHE}^{\rm sym} \frac{(\Delta H)^2 }{\Delta H^2+\small(H-H_{\rm r})^2}+V_{\rm ISHE}^{\rm asym} \frac{2\Delta H(H-H_{\rm r})}{\Delta H^2+\small(H-H_{\rm r})^2},
\end{split}
\end{equation}
where $H$ is the applied magnetic field.
The value of the symmetric part can be taken to be the spin pumping induced ISHE signal ($V_{\rm ISHE}^{\rm sym}$) in our geometry where rectification signals are minimized.~\cite{kumar2018large} 
This is further supported by the fact that signal shape is entirely symmetric and the signal changes sign on reversal of field polarity, both of which are consistent with a dominant ISHE origin of the signal.~\cite{lustikova2015vector} Furthermore, in our geometry the magnetic field is perpendicular to the voltage measurement direction. For this condition, the magnetic spin Hall effect vanishes completely [see Fig. 3 (g) of Ref.\onlinecite{kimata2019magnetic}]. Hence, in the following, we will not consider magnetic spin Hall effect or it's inverse in Mn$_3$Sn.
The charge current generated due to ISHE in Mn$_3$Sn can be written as $V_{\rm ISHE}^{\rm sym} /R$, where $R$ is the resistance of the Mn$_3$Sn/Ni$_{80}$Fe$_{20}$ bilayer. We further normalize the $V_{\rm ISHE}^{\rm sym} /R$ with the width ($w$) of the sample to eliminate any size effect. In order to determine $\SHA$ and $\sdl$ of Mn$_3$Sn we plot $V_{\rm ISHE}^{\rm sym} /Rw$ with $\tMnSn$ (Fig.~\ref{fig:2}(f)) and fit it with the following equation:~\cite{deorani2014observation}

\begin{equation}\label{Vfit}
\begin{split}
\frac{V_{\rm ISHE}^{\rm sym}}{Rw}=\SHA \tMnSn \frac{\hbar\smc\gamma^{2}h_{\rm RF}^{2}(4 \pi M_{s} \gamma + \sqrt{(4 \pi M_{s} \gamma)^2+4\omega^2})}{8 \pi \alphaeff^{2}((4 \pi M_{s} \gamma)^2+4\omega^2)} \\
\times\left(\frac{2e}{\hbar}\right) \frac{\sdl}{\tMnSn} \tanh\left(\frac{\tMnSn}{2\sdl}\right),
\end{split}
\end{equation}

where $h_{\rm RF}$ is the RF field generated due to the RF current of frequency $f=\omega/2\pi$ flowing through the co-planar waveguide, $4\pi \Ms$ is the saturation magnetization, $\hbar$ is the reduced Planck's constant, and $e$ is the electronic charge. From the fitting, we obtain the values of $\sdl$ and $\SHA$ to be 0.42 $\pm$ 0.04 nm and -0.40 $\pm$ 0.03, respectively. The $\smc$ is determined from the enhancement of damping using $\smc=\Delta \alphaeff 4 \pi \Ms t_{\rm Py}/(g\mu_{\rm B})$.~\cite{mosendz2010quantifying} The value is found to be $\smc=1.54\pm0.27\times 10^{19}$ m$^{-2}$ at room temperature. This value of $\smc$ is comparable to the values reported for other systems such as Mn$_3$Ga/CoFeB~\cite{singh2020inverse} and Bi$_2$Se$_3$/Ni$_{80}$Fe$_{20}$~\cite{deorani2014observation} and one order of magnitude higher than Ta/Ni$_{80}$Fe$_{20}$.~\cite{kumar2018large}
The value of $\sdl$ is similar to the value reported for polycrystalline Mn$_3$Sn.~\cite{muduli2019evaluation} The measured $\SHA$ in our case is found to be rather large, especially considering the fact that Eq.~\ref{Vfit} does not include interface transparency, which is less than one. Thus the values of -0.40 $\pm$ 0.03 is a lower limit of $\SHA$. Our value of $\SHA$ is greater than other antiferromagnets, such as $\SHA$~$\approx$~0.31 for polycrystalline Mn$_3$Ga~\cite{singh2020inverse} and $\SHA$~$\approx$~0.35 for (001)-oriented IrMn$_3$.~\cite{zhang2016giant} In addition, the observed value of $\SHA$ is considerably higher than the polycrystalline Mn$_3$Sn value of $\SHA$~$\approx$~0.05, which was reported by one of the authors of this article using a non-local spin transport technique.~\cite{muduli2019evaluation} Our value is also more than twice that of Yu \textit{et al.,}~\cite{yu2021large}, where a value of $\SHA$~$\approx$~$0.18$ was reported for polycrystalline Mn$_3$Sn thin films interfaced with yttrium iron garnet.

In order to compare our results with theoretical calculations~\cite{zhang2018spin} of SHE in Mn$_3$Sn, we determine $\SHC$ using: $\SHC=\SHA \times \sigma \frac{\hbar}{e}$, where $\sigma$ is the charge conductivity of Ru/Mn$_{3}$Sn bilayer, which was found to be 1754 $\Omega ^{-1}$cm$^{-1}$, from four-point probe measurements. We found $\SHC$ to be $-702~\hbar / e$ $\Omega ^{-1}$cm$^{-1}$. 
Zhang \textit{et al.} have predicted an intrinsic SHE in Mn$_3$Sn that arises due to the non-collinear magnetic structure. They predicted $\SHC^{int}\approx 90 (\hbar/e)\Omega ^{-1}$cm$^{-1}$ at the Fermi level. Our value is significantly larger and has the opposite sign, which can be explained if we assume the Fermi level is shifted in our thin films due to slightly higher Mn content. In fact, the band structure is found to be dominated by Mn-$d$ orbitals near the Fermi level~\cite{kuroda2017evidence} and hence, an excess 3\% Mn concentration (Mn$_{3.12}$Sn$_{0.88}$) in our samples can induce electron doping leading to a shift in the Fermi level. Such shift in Fermi level is already reported in Mn$_{3}$Sn both by first principle calculations as well as by angle-resolved photoemission spectroscopy (ARPES) measurements.~\cite{kuroda2017evidence} Based on the measured resistivity and using a scattering time of $62.3$~fs from Ref~\cite{cheng2019terahertz}, we calculated a shift of about 0.09 eV (w.r.t. stoichiometric Mn$_{3}$Sn with no excess Mn) which can easily lead to a sign change in $\SHC$, as shown in the inset of Fig.~\ref{fig:2}(f). However, this can not explain the large magnitude of $\SHC$, indicating that other extrinsic mechanisms may also contribute to $\SHC$. 

In conclusion, we have demonstrated epitaxial growth of $c$-plane oriented non-collinear antiferromagnet Mn$_3$Sn with Ru as a seed layer on Al$_2$O$_3$ substrates. We have investigated ISHE in the $c$-plane Mn$_3$Sn/Ni$_{80}$Fe$_{20}$ system. Through Mn$_3$Sn thickness dependent ISHE measurements, we determine key parameters like $\SHA$ and $\sdl$ for $c$-plane Mn$_3$Sn. We found a large $\SHC$ of $-702~\hbar / e$ $\Omega ^{-1}$cm$^{-1}$, which is higher than other non-collinear antiferromagnets reported till date. The results are important for spin-orbit torque-based spintronics devices utilizing non-collinear antiferromagnets.

\begin{acknowledgments}
The partial support from the Ministry of Human Resource Development under the IMPRINT program (Grant no: 7519 and 7058), the Department of Electronics and Information Technology (DeitY), 
Science \& Engineering research board (SERB File no. CRG/2018/001012), Joint Advanced
Technology Centre at IIT Delhi, and Department of Science and Technology under the Nanomission program (grant no: $SR/NM/NT-1041/2016(G)$) are gratefully acknowledged. H.B. gratefully acknowledges the financial support from the Council of Scientific and Industrial Research (CSIR), Government of India.
\end{acknowledgments}

\section{AUTHOR DECLARATIONS}
\subsection*{Conflict of Interest}
The authors have no conflicts to disclose.

\section{DATA AVAILABILITY}
The data that support the findings of this study are available from the corresponding author upon reasonable request.
\bibliography{Main.bib}

\end{document}